\documentclass[pra,showkeys,superscriptaddress]{revtex4}

\usepackage{bm}
\usepackage{graphicx}
\usepackage{amssymb}
\usepackage{amsmath}
\usepackage{amsthm}
\usepackage{multirow}
\usepackage{mathrsfs}
\DeclareSymbolFont{AMSb}{U}{msb}{m}{n}
\DeclareMathSymbol{\N}{\mathbin}{AMSb}{"4E}
\DeclareMathSymbol{\Z}{\mathbin}{AMSb}{"5A}
\DeclareMathSymbol{\R}{\mathbin}{AMSb}{"52}
\DeclareMathSymbol{\Q}{\mathbin}{AMSb}{"51}
\DeclareMathSymbol{\I}{\mathbin}{AMSb}{"49}
\DeclareMathSymbol{\C}{\mathbin}{AMSb}{"43}

\begin{document}

\title{Classical and quantum fingerprinting with shared randomness and one-sided error}

\author{Rolf T. Horn}
%\email{rhorn@phas.ucalgary.ca}
\affiliation{Institute for Quantum Information Science, University of Calgary, Calgary, Alberta T2N 1N4, Canada}

\author{A. J. Scott}
\email{ascott@qis.ucalgary.ca}
\affiliation{Institute for Quantum Information Science, University of Calgary, Calgary, Alberta T2N 1N4, Canada}

\author{Jonathan Walgate}
\affiliation{Institute for Quantum Information Science, University of Calgary, Calgary, Alberta T2N 1N4, Canada}

\author{Richard Cleve}
\affiliation{Institute for Quantum Information Science, University of Calgary, Calgary, Alberta T2N 1N4, Canada}
\affiliation{Institute for Quantum Computing, University of Waterloo, Waterloo, Ontario N2L 3G1, Canada}

\author{A. I. Lvovsky}
\affiliation{Institute for Quantum Information Science, University of Calgary, Calgary, Alberta T2N 1N4, Canada}

\author{Barry C. Sanders}
\affiliation{Institute for Quantum Information Science, University of Calgary, Calgary, Alberta T2N 1N4, Canada}

\begin{abstract}
Within the simultaneous message passing model of communication complexity,
under a public-coin assumption, we derive the minimum achievable worst-case error
probability of a classical fingerprinting protocol with one-sided error.
We then present entanglement-assisted quantum fingerprinting protocols
attaining worst-case error probabilities that breach this bound.
\end{abstract}

\maketitle

\theoremstyle{plain}
\newtheorem{thm}{Theorem}
\newtheorem{lem}[thm]{Lemma}
\newtheorem{cor}[thm]{Corollary}
\newtheorem{prp}[thm]{Proposition}
\newtheorem{con}[thm]{Conjecture}

\theoremstyle{definition}
\newtheorem{dfn}[thm]{Definition}

\theoremstyle{remark}
\newtheorem*{rmk}{Remark}
\newtheorem*{pprf}{Proof}
\newtheorem{exm}{Example}

\def\tr{\operatorname{tr}}
\def\sgn{\operatorname{sgn}}
\def\ket#1{|#1\rangle}
\def\bra#1{\langle#1|}
\def\inner#1#2{\langle #1 | #2 \rangle}

\def\Pe{P_\text{\rm e}}
\def\Pwce{P_\text{\rm wce}}
\def\Ne{N_\text{\rm e}}
\def\sr{\sigma}

\section{Introduction}

Computing whether two binary strings are equal or not is an important
task that can be used to protect software, or used as a primitive for authentication.
Unfortunately the comparison of two objects, such as two operating systems,
may be expensive when the entire message strings that identify these objects must be
transmitted over large distances. Fingerprinting allows a significant reduction in
communication costs when a small likelihood of error in the comparison is acceptable.
Then, rather than transmitting the entire message string for the object itself, a relatively
shorter string, or fingerprint, that identifies the object is sent. Although errors may
arise in the comparison of fingerprints, this error can be made sufficiently small
by simply increasing the fingerprint length.

The key question concerned with fingerprinting is, for given message and fingerprint lengths,
what is the minimum achievable guaranteed error rate? In this article we partially answer this
question for fingerprinting protocols described within Yao's simultaneous message passing
model of communication complexity~\cite{Yao79,Kushilevitz97}. The fingerprints are then
generated and transmitted by two parties, Alice and Bob, who are forbidden direct communication, 
but instead allowed to correspond with a referee known as Roger.

\begin{figure}[t]
\includegraphics[scale=0.5]{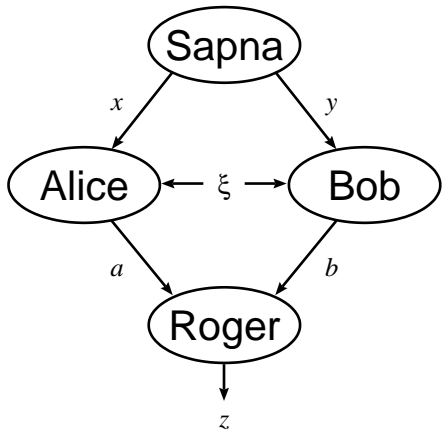}
\caption{The communication flow diagram for fingerprinting with a shared random key $\xi$.} 
\label{fig}
\end{figure}

Our fingerprinting scenario is described as follows (see Fig \ref{fig}). A supplier, who we 
call Sapna, chooses two messages, $x$ and $y$, from a pool of $n$ unique messages and hands them 
to Alice and Bob, respectively. As communication is considered expensive,
Alice and Bob are limited to sending fingerprints of their original
messages to Roger, $a$ and $b$ respectively, which they select from a smaller
pool of size $m$. Roger then infers
\begin{equation}
\label{Eq}
    \text{EQ}(x,y)
        =\left\{\begin{array}{ll}1,&\text{if }x=y\\ 0,&\text{if }x\neq y\end{array}\right.\;,
\end{equation}
and completes the protocol by revealing a single bit $z\in\{0,1\}$.
Roger is correct if $z=\text{EQ}(x,y)$. In the current investigation we consider
one-sided-error protocols, in which case, $z=0$ only if $x\neq y$. One-sided-error protocols are of vital practical importance whenever the `cost' of false negative results exceeds that of false positives.

The fingerprinting protocol adopted by Alice, Bob and Roger is publicly announced. The goal
of this protocol is to minimize Roger's error probability. Sapna, however, may be a saboteur,
and always choose message pairs that lead to the highest rate of error in Roger's output.
We thus evaluate fingerprinting protocols according to this worst-case scenario. The
worst-case error probability, $\Pwce=\max_{x,y}\text{Pr}(z\neq\text{EQ}(x,y))$, then
corresponds to the maximum error rate of the protocol.

In the private-coin model, each party is handed a coin to generate private randomness.
This gives Alice and Bob the ability to probabilistically avoid message collisions,
in which different messages produce the same fingerprint.
In the following we analyze the public-coin model, for which an additional source
of randomness is made available, in the form of a secret key generated by a public coin, to
be shared by Alice and Bob, but kept hidden from Sapna. One way to hide the key
from Sapna is for Alice and Bob to use only those public-coin outcomes that have
arisen after Sapna has dealt the messages.

There has been recent interest in quantum analogues of fingerprinting protocols \cite{Buh01,Niel04,Yao03,Ambainis03,Gavinsky04,Nishimura04}.
Whereas classical fingerprints are length $\log_2 m$ bit strings, quantum fingerprints
are states in an $m$-dimensional Hilbert space, or equivalently, $\log_2 m$ qubit strings.
Furthermore, in the quantum regime, shared randomness is replaced by
shared entangled states. The seemingly more general case of Alice and Bob sharing both
entanglement and randomness is not necessary: Alice and Bob may always generate
shared randomness from shared entanglement through local measurements.

It has been shown in the asymptotic limit that, when shared randomness (or entanglement) is
forbidden, fingerprints composed of quantum information can be made exponentially smaller than
those composed of classical information. Specifically, for messages of length $N\equiv\log_2 n$ bits,
in the classical case it is sufficient and necessary for Alice and Bob to communicate fingerprints
of length $O(\sqrt{N})$ bits if the error is to be kept arbitrarily small \cite{Ambainis96,Newman96,Babai97}.
If however, the parties communicate fingerprints constructed from quantum bits, only $O(\log N)$ many are
needed \cite{Buh01}. This definitive resource advantage does not exist when a shared key is allowed,
in which case, fingerprints of length $O(1)$ bits/qubits are now sufficient \cite{Yao79,Buh01,Babai97}.
Here we derive an analytic bound, however, that quantum fingerprinting protocols must surpass in order to
claim any advantage over classical protocols. Such bounds are important for experimental tests of
quantum fingerprinting \cite{Horn04,Du04}.

We show that, for classical fingerprinting protocols with one-sided error and
an arbitrary amount of shared randomness, the \emph{minimum achievable} worst-case
error probability is
\begin{equation}
\label{eq:Pclassical_wce}
    \frac{k \lceil n/m \rceil^2 + (m-k)\lfloor n/m \rfloor^2 - n}{n^2-n}
\end{equation}
($k= n\mod m$) when $n\geq m$, and $0$ otherwise. Quantum
fingerprinting protocols with an arbitrary amount of shared
entanglement, on the other hand, are shown to attain worst-case
error probabilities of 
\begin{equation}
\label{eq:Pquantum_wce}
    \frac{n/m^2-1}{n-1}
\end{equation}
when $n\geq m^2$, and $0$ otherwise. The difference between the two error rates is made clear when 
$m$ divides $n$, in which case the classical error probability reduces to $(n/m-1)/(n-1)$.
It is interesting that the addition of shared entanglement in the quantum case allows perfect error-free
fingerprinting protocols to be constructed when $m<n$. 
In the limit of large message numbers, $n\rightarrow\infty$, the classical error
probability~(\ref{eq:Pclassical_wce}) tends to $1/m$ whereas the quantum error
probability~(\ref{eq:Pquantum_wce}) approaches $1/m^2$. Thus, in the asymptotic limit,
some improvement of quantum fingerprinting protocols over classical protocols still exists
in the presence of shared randomness or entanglement. We now begin our analysis by 
considering classical fingerprinting protocols.

\section{Classical strategies with shared randomness}
\label{sec:classical}

We first present a simple protocol which
achieves the bound [Eq.~(\ref{eq:Pclassical_wce})]. In each round of fingerprinting, 
Alice and Bob use their shared random key to partition
the set of $n$ messages into $m$ groups of almost equal
size: $k$ groups containing $\lceil n/m \rceil$ messages, and $m-k$
groups containing $\lfloor n/m \rfloor$ messages ($k= n\mod m$). This partition
is identical for Alice and Bob, but given the randomness of the key,
is completely unknown to Sapna. Upon receipt of the messages, Alice 
and Bob generate fingerprints according to which group they belong to. 
Roger then infers equality if and only if the fingerprints he receives are 
identical i.e. the messages belong to the same group. 

In this protocol, the worst-case scenario occurs when Sapna chooses 
unequal message pairs belonging to the same group. Sapna has 
$k \lceil n/m \rceil^2 + (m-k)\lfloor n/m \rfloor^2-n$ choices 
from a total of $n^2-n$ unequal message pairs, but not being privy to 
how the messages are grouped, she is instead compelled to send 
random pairs of unequal messages. The worst-case error rate is thus 
given by the ratio of these two numbers [Eq.~(\ref{eq:Pclassical_wce})].
Note that the protocol is implemented without any need for private 
randomness. The remainder of this section is dedicated to proving that it
is indeed optimal. 

It will prove useful to think of Alice, Bob and Roger as a team with the
shared goal of maximizing Roger's probability of success, and Sapna operating as their
opponent. This team has a pre-established, publicly known strategy.
In this strategy, Alice and Bob have a probability of communicating each
fingerprint pair $(a,b)$ to Roger for a given message pair $(x,y)$ provided by Sapna.
Furthermore, in this strategy, Roger has a fixed probability of declaring
$x$ and $y$ to be the same message upon receipt of fingerprint pair $(a,b)$ provided by Alice and Bob.
Any strategy is completely specified by a triple of functions $(p,q,r)$, where
$p,q:\{1, \ldots ,m\}\times\{1, \ldots ,n\}\times\mathbb{Z}^+ \rightarrow [0,1]$ and
$r:\{1, \ldots ,m\}\times\{1, \ldots ,m\}\rightarrow [0,1]$. The function $p(a|x,\xi)$ is
the probability that Alice sends fingerprint $a$ to Roger, given that she receives
message $x$ from Sapna and shares the random key $\xi$ with Bob. Similarly, $q(b|y,\xi)$ is the
probability that Bob sends $b$ to Roger, given that he receives $y$
from Sapna and shares $\xi$ with Alice. The function $r(a,b)$ is the
probability that Roger outputs $z=1$, given that he receives fingerprint $a$
from Alice and $b$ from Bob.

When a party's private strategy ($p$, $q$ or $r$) takes values
only in the set $\{0,1\}$, we call that party's strategy {\it
deterministic}. If all parties' strategies are deterministic we
call the triple $(p,q,r)$ a \emph{deterministic strategy}.
Otherwise a general (i.e. probabilistic) strategy should be
assumed. Normalization requires
\begin{equation}
\sum_{a=1}^{m}p(a|x,\xi)=\sum_{b=1}^{m}q(b|y,\xi)=1
\label{pqnorm}\end{equation}
for all $x,y $ and $\xi$.

Our source of shared randomness is expressed through the function
$\sr:\mathbb{Z}^+\rightarrow [0,1]$, where $\sr(\xi)$ is the
probability that Alice and Bob share the state $\xi$, and normalization requires
\begin{equation}
\sum_{\xi=1}^\infty\sr(\xi)=1 \;.
\end{equation}
To obtain absolute bounds on the performance of classical
strategies, we allow Alice and Bob to share arbitrarily large
amounts of randomness, or equivalently, we allow Alice and Bob to
choose $\sr$. The triple $(p,q,r)$ is then referred to as a {\it
strategy with shared randomness}. If, however, Alice and Bob are instead
constrained to use a particular distribution, $\sr$, we will
call the triple $(p,q,r)$ a \emph{strategy with shared
randomness $\sr$}. Finally, we call $(p,q,r)$ a
\emph{strategy without shared randomness} whenever both Alice and
Bob use strategies that are independent of $\xi$.

Given a strategy $(p,q,r)$ with shared randomness $\sr$, the
probability that Roger outputs 1 when Sapna deals $x$ to Alice and
$y$ to Bob is
\begin{equation}
P_1^{(p,q,r)}(x,y)\equiv\sum_{\xi=1}^\infty \sum_{a,b=1}^m p(a|x,\xi)q(b|y,\xi)r(a,b)\sr(\xi)\;.
\label{p1eq}\end{equation}
Defining the \emph{error probability}
\begin{equation}
\Pe^{(p,q,r)}(x,y)\equiv\left\{\begin{array}{ll}
1-P_1^{(p,q,r)}(x,x), & x=y \\
P_1^{(p,q,r)}(x,y), & x\neq y\end{array}\right.\;,
\end{equation}
the \emph{worst-case error probability} is then simply the largest error
probability that Sapna can coerce
\begin{equation}
\Pwce^{(p,q,r)}\equiv\max_{x,y}\;\Pe^{(p,q,r)}(x,y)\;,
\end{equation}
and an \emph{optimal strategy} is one that results in the smallest
possible worst-case error probability, solving the minimax problem
\begin{equation}
\min_{p,q,r}\;\max_{x,y}\;\Pe^{(p,q,r)}(x,y)=\min_{p,q,r}\;\Pwce^{(p,q,r)}\;.
\label{minimax}\end{equation}
A strategy is said to have \emph{one-sided error} when
\begin{equation}
P_1^{(p,q,r)}(x,x)=1
\label{onesideconstraint}\end{equation}
for all $x$. Using such a strategy, it is impossible for Roger to announce 0
when Sapna has supplied Alice and Bob with identical messages. For the current
investigation we consider only one-sided-error strategies.

To begin, let us introduce a lemma that allows the following simplification.
Whereas Roger can use a probabilistic strategy~$r$, we show that there exists a
deterministic strategy~$r'$ for Roger that is at least as good as all probabilistic 
strategies.
\begin{lem}
Let $(p,q,r)$ be a fingerprinting strategy with shared randomness
$\sr$ and one-sided error. Then
\begin{equation}
\Pe^{(p,q,r)}(x,y)\;\geq\;\Pe^{(p,q,r')}(x,y)
\end{equation}
for all $x$ and $y$, where
\begin{equation}
    r'(a,b)=\left\{\begin{array}{ll}
    1,&\text{if $p(a|x,\xi)>0$ and $q(b|x,\xi)>0$ for some $x$ and $\xi$}\\
    0,&\text{otherwise}\end{array}\right.\;.
\label{roger}\end{equation}
\label{lemma1}\end{lem}
\begin{proof}
Given a particular $p$ and $q$,
to satisfy the one-sided-error constraint [Eq.~(\ref{onesideconstraint})] we must necessarily have
$r(a,b)=1$ whenever there is an $x$ and $\xi$ such that
$p(a|x,\xi)>0$ and $q(b|x,\xi)>0$. Our goal is to now minimize
$P_1(x,y)$ whenever $x\neq y$, and thus, setting $r(a,b)=0$ in the
remaining cases is optimal.
\end{proof}

Lemma \ref{lemma1} allows us to limit our search for optimal one-sided-error
strategies to the class where Roger's decisions are given by Eq.~(\ref{roger}),
and are thus purely deterministic. Define the quantity
\begin{equation}
\Ne^{(p,q,r)} \equiv \sum_{x,y} \Pe^{(p,q,r)}(x,y). \label{N}
\end{equation}
This quantity, for deterministic one-sided-error strategies with no shared randomness, is the total number of message pairs $(x,y)$ that
produce an error. For more general strategies the quantity $\Ne^{(p,q,r)}$ can be used to derive bounds on the worst-case error probability.

\begin{lem}
Let $(p,q,r)$ be a fingerprinting strategy with shared randomness
$\sr$ and one-sided error. Then there exists a deterministic
fingerprinting strategy, $(p',q',r')$, without shared randomness but
with one-sided error, such that
\begin{equation}
\Ne^{(p,q,r)} \;\geq\; \Ne^{(p',q',r')}\;. \label{preboundN}
\end{equation}
\label{lemma2}\end{lem}
\begin{proof}
First replace Roger's strategy, $r$, by the deterministic strategy, 
$r'$ [Eq.~(\ref{roger})]. Then by Lemma \ref{lemma1}, 
\begin{equation}
\Ne^{(p,q,r)}\geq \Ne^{(p,q,r')}\;.
\label{Neq1}\end{equation}

Now define the strategies without shared randomness, $(p_\xi,q_\xi,r')$, by setting 
$p_\xi(a|x)\equiv p(a|x,\xi)$ and $q_\xi(a|x)\equiv q(a|x,\xi)$ for each $\xi$. Then
\begin{equation}
    \Ne^{(p,q,r')} =\sum_{\xi}\Ne^{(p_\xi,q_\xi,r')}
    \sr(\xi) \geq \min_{\xi}\Ne^{(p_\xi,q_\xi,r')}=\Ne^{(p_{\xi'},q_{\xi'},r')}\;,
\label{Neq2}\end{equation}
where $(p_{\xi'},q_{\xi'},r')$ is a strategy without shared randomness which achieves the minimum.

The functions $p_{\xi'}$ and $q_{\xi'}$ are probabilistic private strategies without shared randomness. 
Under the normalization constraint [Eq.~(\ref{pqnorm})], the set of all such private strategies 
is convex and compact. The extreme points of this set are precisely the $m^n$ different deterministic 
strategies. Since any member of a compact convex set can be rewritten in terms of a convex combination 
of the extreme points, any probabilistic strategy can be rewritten in terms of a convex combination of 
deterministic strategies. Specifically, we can rewrite Alice's and Bob's strategies as
\begin{equation}
p_{\xi'}(a|x) = \sum_i\phi_i \hat{p}_i(a|x) \qquad\text{ and }\qquad q_{\xi'}(b|y) = \sum_j\theta_j \hat{q}_j(b|y) \;,
\end{equation}
respectively, where $\phi_i,\theta_j\geq 0$, $\sum_j\phi_j=\sum_j\theta_j=1$, and the strategies
$\hat{p}_i(a|x)$ and $\hat{q}_j(b|y)$ are deterministic. Alice and Bob may now enact
the strategy $(p_{\xi'},q_{\xi'},r')$ by each flipping private coins to determine which 
$i$ and $j$ to use before Sapna deals $(x,y)$. The probability that they choose pair $(i,j)$ is 
then $\phi_i\theta_j$, and
\begin{equation}
\Ne^{(p_{\xi'},q_{\xi'},r')} = \sum_{i,j}\phi_i\theta_j \Ne^{(\hat{p}_i,\hat{q}_j,r')}\geq \min_{i,j} \Ne^{(\hat{p}_i,\hat{q}_j,r')}=\Ne^{(p',q',r')}\;,
\label{Neq3}\end{equation}
where $(p',q',r')$ is a deterministic strategy that achieves the minimum.
Combining inequalities (\ref{Neq1}), (\ref{Neq2}) and (\ref{Neq3}) completes the proof.
\end{proof}

Lemma \ref{lemma2} implies that neither private nor shared randomness is needed for the minimization of $\Ne^{(p,q,r)}$.
A deterministic fingerprinting strategy without shared randomness will suffice. In the following lemma
we give such a strategy.

\begin{lem}
Let $(p,q,r)$ be a deterministic fingerprinting strategy without shared randomness but with one-sided error. Then
\begin{equation}
    \Ne^{(p,q,r)} \;\geq\; k \lceil n/m \rceil^2 + (m-k)\lfloor n/m \rfloor^2 - n
\label{boundN}
\end{equation}
where $k= n\mod m$. Furthermore, equality holds for the strategy with
\begin{equation}
    r(a,b)=\delta(a,b) \qquad\text{ and }\qquad p(a|x)=q(a|x)=\left\{\begin{array}{ll}
    1,&\text{if $a-1\equiv x-1 \mod m$}\\
    0,&\text{otherwise}\end{array}\right..
\label{optimal}\end{equation}
\label{lemma3}\end{lem}

\begin{proof}
By Lemma \ref{lemma1}, under the one-sided error condition it is optimal for Roger to employ the 
deterministic strategy given by Eq.~(\ref{roger}). Assume this to be the case for the remainder of the proof.

Suppose Alice and Bob also employ deterministic strategies; they translate every incoming message to a specific fingerprint.
Their joint strategy may be described by a pair of many-to-one maps drawn from the set of $m^n$ different fingerprinting functions of the form
$f:\{1, \ldots ,n\}\rightarrow\{1, \ldots ,m\}$. Specifically, $p(a|x)\equiv\delta(f^{(p)}(x),a)$ and
$q(b|y)=\delta(f^{(q)}(y),b)$ where $f^{(p)}$ and $f^{(q)}$ are Alice's and Bob's
fingerprinting functions, respectively. Roger's strategy is thus
\begin{equation}
r(a,b)=\left\{\begin{array}{ll}
1,&\text{if $f^{(p)}(x)=a$ and $f^{(q)}(x)=b$ for some $x$}\\
0,&\text{otherwise}\end{array}\right. \;.
\end{equation}
Define the message sets
\begin{equation}
	M_a^{(p)}\equiv\left\{x\,|\,f^{(p)}(x)=a\right\}\;, \qquad M_b^{(q)}\equiv\left\{y\,|\,f^{(q)}(y)=b\right\}\;,
\end{equation}
which contain all messages mapped to Alice's fingerprint, $a$, and Bob's fingerprint, $b$, respectively.
The quantity
\begin{equation}
	s_{ab}\equiv\left|M_a^{(p)}\cap M_b^{(q)}\right| ,
\end{equation}
counts the number of equal message pairs $(x,x)$ mapped to fingerprint pair $(a,b)$, and likewise
\begin{equation}
d_{ab}\equiv\left|M_a^{(p)}\right|\cdot\left|M_b^{(q)}\right|-s_{ab}
\end{equation}
is the number of unequal message pairs $(x,y)$ mapped to fingerprint pair $(a,b)$. Notice that,
since both $\{M_a^{(p)}\}_{a=1}^m$ and $\{M_b^{(q)}\}_{b=1}^m$ form set partitions of $\{1, \ldots ,n\}$,
we have the following relations
\begin{equation}
\sum_a s_{ab} = \left|M_b^{(q)}\right|\;, \qquad \sum_b s_{ab} = \left|M_a^{(p)}\right| \;, \qquad \sum_{a,b} s_{ab} = n \;,
\end{equation}
and hence
\begin{equation}
d_{ab}=\Bigg(\sum_{i,j} s_{ai}s_{jb}\Bigg)-s_{ab}\;.
\end{equation}
The total number of message pairs $(x,y)$ that produce an error is then
\begin{equation}
\Ne^{(p,q,r)} = \sum_{a,b} d_{ab}r(a,b) = \Bigg(\sum_{a,b,i,j} s_{ai}s_{jb}\sgn(s_{ab})\Bigg) - n \equiv F(s)-n \;,
\end{equation}
where Roger's strategy is now expressed as $r(a,b)=\sgn(s_{ab})$ to emphasize the explicit dependence on the matrix $s$. 
The convention $\sgn(0)=0$ is used for the signum function.

We now minimize $F(s)$ over all $m\times m$ matrices $s$ with nonnegative integer entries, subject to the constraint $\sum_{a,b} s_{ab} = n$.  
First note that we may assume $s$ is diagonal. If it were not, we could define the diagonal matrix $s'$ with nonzero entries
$s'_{aa}=\sum_j s_{aj}$ and the property
\begin{eqnarray}
F(s') = \sum_{a,b,i,j} s'_{ai}s'_{jb}\sgn(s'_{ab}) &=& \sum_{a} {s'_{aa}}^2 \\
&=& \sum_{a,b,i} s_{ai}s_{ab} \\
&=& \sum_{a,b,i} s_{ai}s_{ab}\sgn(s_{ab}) \\
&\leq& \sum_{a,b,i,j} s_{ai}s_{jb}\sgn(s_{ab}) = F(s)\;.
\end{eqnarray}
For $s$ diagonal, the minimum of $F(s)=\sum_a {s_{aa}}^2$ under the constraint $\sum_a s_{aa}=n$ clearly occurs when
$s_{aa}= \lceil n/m \rceil$ for $k$ entries, and $s_{aa}= \lfloor n/m \rfloor$ for $m-k$ entries, 
and thus, the number of message pairs which produce an error is bounded below by the RHS of Eq.~(\ref{boundN}).
To complete the proof it is trivial to check that the inequality saturates under the given strategy 
[Eq.~(\ref{optimal})].
\end{proof}

The above three lemmas allow us to prove our main result in a straightforward fashion.

\begin{thm}
Let $(p,q,r)$ be a fingerprinting strategy with shared randomness and one-sided error. Then
\begin{equation}
\Pwce^{(p,q,r)}\geq\frac{k \lceil n/m \rceil^2 + (m-k)\lfloor n/m \rfloor^2 - n}{n^2-n}
\label{boundWCE}\end{equation}
where $k= n\mod m$. Furthermore, equality
holds when Alice and Bob use the deterministic strategy of Lemma \ref{lemma3}
after applying a completely random permutation to the labels of Sapna's messages through the shared
randomness. That is, they use the strategy with
\begin{equation}
    r(a,b)=\delta(a,b) \qquad\text{ and }\qquad p(a|x,\xi)=q(a|x,\xi)
    =\left\{\begin{array}{ll}
    1,&\text{if $\pi_\xi(a)-1\equiv x-1 \mod m$}\\
    0,&\text{otherwise}\end{array}\right.
\label{optimal2}\end{equation}
where $\pi_\xi$ is one of $n!$ different permutations of Sapna's message labels,
and $\sr(\xi)=1/n!$ for $1\leq\xi\leq n!$ (and zero otherwise), is chosen for the shared randomness.
\label{theorem4}\end{thm}
\begin{proof}
From Eq.~(\ref{N}) the average error probability of a one-sided-error strategy, taken over all unequal
message pairs, is given by $\Ne^{(p,q,r)}/(n^2-n)$. This average error probability provides
a lower bound for the worst-case error probability. Thus, by Lemmas \ref{lemma2} and \ref{lemma3},
we have
\begin{equation}
\Pwce^{(p,q,r)}\geq\frac{\Ne^{(p,q,r)}}{n^2-n}\geq\frac{k \lceil n/m \rceil^2 + (m-k)\lfloor n/m \rfloor^2 - n}{n^2-n} \;.
\end{equation}
The first inequality saturates if Alice and Bob apply a random permutation to Sapna's message labels
immediately after $x$ and $y$ are dealt; the second saturates if they follow this permutation
by the deterministic strategy of Lemma~\ref{lemma3} [Eq. \ref{optimal}].
\end{proof}

Note that no private randomness is needed for the optimal strategy. In all of the above we have assumed
that Alice and Bob are the only parties allowed access to the random source $\sr$. When we also grant Roger access, replacing
$r(a,b)$ by $r(a,b,\xi)$, straightforward adjustments to the above proof show that Eq.~(\ref{boundWCE}) again
applies. If however, Sapna is also granted access, it is obvious that our fingerprinting scenario will
revert to one without shared randomness. Note that if the value of $\xi$ is announced publicly at set
intervals, Alice and Bob may always deny Sapna knowledge of $\xi$, by simply using only those values
announced after $x$ and $y$ are dealt.

We can investigate the classical
communication complexity of fingerprinting with shared randomness by considering cases where equality
holds in Eq.~(\ref{boundWCE}). Then $\Pwce<1/m$, and consequently, $\log_2(1/\epsilon)=O(1)$ fingerprint
bits are sufficient to keep $\Pwce<\epsilon$ for any small fixed $\epsilon>0$.
Defining the number of message and fingerprint bits, $N\equiv\log_2(n)$ and $M\equiv\log_2(m)$, respectively,
we see that the above optimal protocol [Eq.~(\ref{optimal2})] requires $\log_2(n!)=O(2^NN)$ bits of shared
randomness. By discarding repetitions in the set of $n!$ deterministic strategies implicit in
Eq.~(\ref{optimal2}), we can reduce this to $\log_2\!\big(n!/[(\lceil n/m \rceil !)^k(\lfloor n/m \rfloor !)^{m-k}(m-k)!k!]\big)=O\big((2^N-2^M)M\big)$ bits of shared randomness, but this is still
hugely excessive. If we relax the condition of strict optimality to strategies which simply keep the number of
fingerprint bits $O(1)$ in message size, and the error arbitrarily small, only $O(\log(N))$ bits of shared randomness
will suffice \cite{Newman91,Babai97,Buh01}.

Finally, we remark that if Bob is given a larger set of fingerprints, the minimum achievable worst-case
error probability remains the same. In fact, in the general case where Alice has $m_A$ fingerprints and Bob 
has $m_B$ fingerprints, Theorem \ref{theorem4} applies if we set $m=\min\{m_A,m_B\}$ throughout.  
We can show this as follows. First note that Lemma \ref{lemma1} and \ref{lemma2} are unaffected by the generalization.
To generalize Lemma \ref{lemma3} we need only consider the special case where $m_A=m\leq n=m_B$. For deterministic 
strategies with $m_B=n$, without loss of generality, we may set $q(b|y)=\delta_{by}$ so that Bob simply passes on Sapna's message to 
Roger. Lemma \ref{lemma1} then implies that the optimal choice for Roger's strategy is $r(a,y)=p(a|y)$. The total
resulting strategy $(p,q,r)$, however, is now equivalent to the strategy $(p',q',r')$, where $p'(a|x)\equiv q'(a|x)\equiv p(a|x)$ 
and $r'(a,b)\equiv\delta_{ab}$, in that $\Pe^{(p,q,r)}(x,y)=\Pe^{(p',q',r')}(x,y)$ for all $x$ and $y$. 
Note that the strategy $(p',q',r')$ makes no use of Bob's additional fingerprints $m<b\leq m_B$, and hence, we have shown
that it is possible to convert deterministic strategies with the parameters $m_A=m\leq n=m_B$ to those with $m_A=m_B=m$ without 
changing the error rate. Consequently, Lemma \ref{lemma3} must also apply to the special case 
$m_A=m\leq n=m_B$, and given that the minimum possible value of $\Ne^{(p,q,r)}$ cannot decrease when $m_B$ is decreased, 
Lemma \ref{lemma3} applies to the general fingerprinting scenario if we set $m=\min\{m_A,m_B\}$ throughout. 
Theorem \ref{theorem4} now follows but with all cases of $m$ replaced by $\min\{m_A,m_B\}$.

\section{Quantum strategies with shared entanglement}

In the quantum scenario we replace Alice's and Bob's classical fingerprints ($a$ and $b$) and probability 
distributions [$p(a|x,\xi)$ and $q(b|y,\xi)$], by quantum states, $\hat{\rho}(x,\hat{\sr})$ and $\hat{\tau}(y,\hat{\sr})$ respectively, 
of an $m$-dimensional Hilbert space, denoted by $\mathcal{H}_m$,
and the shared randomness $\sr(\xi)$ by an entangled quantum state $\hat{\sr}$ of the tensor-product space
$\mathcal{H}_{d_A}\otimes\mathcal{H}_{d_B}$, where $\mathcal{H}_{d_A}$ belongs to Alice and
$\mathcal{H}_{d_B}$ to Bob. In the following analysis, all such quantum states will be pure.
In correspondence with the classical scenario, we can either restrict Alice and Bob to use a particular
given $\hat{\sr}$, calling a protocol satisfying this constraint a
{\it strategy with shared entanglement $\hat{\sr}$}, or grant them any choice of entangled state, in
which case we simply say the protocol is a {\it strategy with shared entanglement}. Being a pre-established
component of the fingerprinting apparatus, Sapna will be allowed
knowledge of $\hat{\sr}$, just as she is allowed knowledge of the probability distribution $\sr(\xi)$ in the
classical scenario. For the tensor-product space $\mathcal{H}_d\otimes\mathcal{H}_d$ ($d_A=d_B=d$), 
define the maximally entangled quantum state
$\ket{\psi_+^{(d)}}\equiv d^{-1/2}\sum_{k=1}^d \ket{k}_A\otimes\ket{k}_B$, where 
$\ket{k}_A$ and $\ket{k}_B$ are basis states for Alice and Bob respectively. In the following, Alice 
and Bob use the same computational basis, in which case we drop the subscripts.
Our first result shows that whenever $n\leq m^2$ error-free quantum fingerprinting
strategies exist.

\begin{thm}
When $n \leq m^2$ there exists an error-free quantum fingerprinting strategy with shared entanglement
$\hat{\sr}=\ket{\psi_+^{(m)}}\bra{\psi_+^{(m)}}$.
\label{theorem5}\end{thm}
\begin{proof}
Let $\{U_x\}_{x=1}^{m^2}$ be an orthonormal unitary operator basis for $\text{End}(\mathcal{H}_m)$, the space 
of linear operators acting on $\mathcal{H}_m$ i.e. $\tr\left[U_x^{\dag}U_y\right] = m\,\delta_{xy}$. For example, we could use the operators 
defined by Eq. (\ref{UTOF}) below with $n=m^2$. 

Upon receipt of Sapna's messages $x$ and $y$, Alice and Bob perform on their portions of $\ket{\psi_+^{(m)}}$
the unitaries $U_x^*$ and $U_y$, respectively, where conjugation is done in the computational basis,
and pass the resulting state on to Roger. Noting that
\begin{equation}
\bra{\psi_+^{(m)}}U_x^*\otimes U_y\ket{\psi_+^{(m)}} =
\frac{1}{m}\sum_{j,k} \bra{k}U_x\ket{j}^*\bra{k}U_y\ket{j} =
\frac{1}{m}\sum_{j,k} \bra{j}U_x^\dag\ket{k}\bra{k}U_y\ket{j} =
\frac{1}{m}\tr\left[U_x^{\dag}U_y\right] = \delta_{xy}
\end{equation}
we find that the state received by Roger remains equal to
$\ket{\psi_+^{(m)}}$ when $x=y$, and orthogonal to $\ket{\psi_+^{(m)}}$ when
$x\neq y$. With the projective measurement
$\big\{P_1=\ket{\psi_+^{(m)}}\bra{\psi_+^{(m)}}, P_0=1-P_1\big\}$,
Roger faultlessly determines ${\rm EQ}(x,y)$.
\end{proof}

Notice that without classical communication, Alice and Bob cannot convert $\log_2 m$ (or more) entangled qubits into
the maximally entangled quantum state, $\ket{\psi_+^{(m)}}$ (but both quantities can be converted into $\log_2 m$
privately shared random bits). In the classical case, however, Alice and Bob can convert $\sigma$ into
approximately $\sum_\xi\sigma(\xi)\log_2\sigma(\xi)$ uniformly random bits, and vice versa, by simply
agreeing to a pre-established formula. Thus shared randomness is an interconvertible resource, whereas
shared entanglement is not.

The quantum fingerprinting protocol used for the proof of Theorem \ref{theorem5} 
may be extended to cases where $n>m^2$ by means of a straightforward reformulation 
of the classical strategy described in the beginning of Section~\ref{sec:classical}, 
with the number of groups now being $m^2$ rather than $m$. The error rate of this protocol 
is given by
\begin{equation} \label{eq:Psemiclassical_wce}
    \Pwce=\frac{k \lceil n/m^2 \rceil^2 + (m^2-k)\lfloor n/m^2 \rfloor^2 -n}{n^2-n}\;,
\end{equation}
where $k= n\mod m^2$.

An improved error rate can be achieved using the
following approach. For each $\epsilon>0$ we evaluate how many
unitary operators $U_x$ we can construct with the property
$\sum_{x,y}\left|\tr\left[U_x^{\dag}U_y\right]\right|^2\leq\epsilon$.
It can be shown that 
\begin{equation}
\sum_{x,y=1}^n |\tr(E_x^\dag E_y)|^2 \geq n^2
\label{framebound}\end{equation}
for any set $\{E_x\}_{x=1}^n\subset\text{End}(\mathcal{H}_m)$ of $n\geq
m^2$ linear operators with normalization $\tr(E_x^\dag E_x)=m$ for
all $x$ \cite{welch,Benedetto03}. The proof of the following theorem
relies on the existence of a set of {\it unitary} operators
achieving this bound. Note that when $n=lm^2$, where $l$ is a positive integer, 
the error rates of Eq. (\ref{eq:Psemiclassical_wce}) and Eq. (\ref{Qwcep}) 
below coincide. This is a consequence of the fact that $l$ copies of an 
orthonormal unitary operator basis will saturate the inequality [Eq. (\ref{framebound})]. 

\begin{thm}
When $n \geq m^2$ there exists a quantum fingerprinting strategy with shared entanglement 
$\hat{\sigma}=\ket{\psi_+^{(m)}}\bra{\psi_+^{(m)}}\otimes\ket{\psi_+^{(n!)}}\bra{\psi_+^{(n!)}}$, and worst-case error probability
\begin{equation}
\Pwce=\frac{n/m^2-1}{n-1}\;.
\label{Qwcep}\end{equation}
\label{theorem6}\end{thm}
\begin{proof}
For $n\geq m^2$ define the set $\{U_x\}_{x=1}^n\subset\text{End}(\mathcal{H}_m)$ of unitary operators with matrix components
\begin{equation}
\bra{j}U_x\ket{k}\;\equiv\; \frac{1}{\sqrt{m}}\exp\left[\frac{2\pi ijk}{m}+\frac{2\pi i(j+mk)x}{n}\right]\;,
\label{UTOF}\end{equation}
where now $i\equiv\sqrt{-1}$. When $n=m^2$, $\{U_x\}_{x=1}^{m^2}$ forms an orthonormal unitary operator basis, and in general, a {\it tight unitary
operator frame} \cite{ScottXX}. It is simple to verify unitarity of the operators,
\begin{equation}
\bra{j}U_x^\dag U_x\ket{k}\;=\;\sum_{l=1}^m{\bra{l}U_x\ket{j}}^*\bra{l}U_x\ket{k}\;=\;\frac{1}{m}\sum_{l=1}^m\exp\left[\frac{2\pi i (k-j)l}{m}+\frac{2\pi im(k-j)x}{n}\right]\;=\;\delta_{jk}\;,
\end{equation}
orthogonality when $n=m^2$,
\begin{eqnarray}
\tr\left[U_x^\dag U_y\right] &=& \sum_{j,k=1}^m{\bra{j}U_x\ket{k}}^*\bra{j}U_y\ket{k} \\
&=& \frac{1}{m}\sum_{j,k=1}^m\exp\left[\frac{2\pi i(j+mk)(x-y)}{n}\right] \\ 
&=& \frac{1}{m}\sum_{l=1}^{m^2}\exp\left[\frac{2\pi i(l+m)(x-y)}{m^2}\right] \\
&=& m\,\delta_{xy}\;,
\end{eqnarray}
and that
\begin{eqnarray}
\sum_{x,y=1}^n\left|\tr\left[U_x^\dag U_y\right]\right|^2 &=& \sum_{x,y=1}^n\sum_{j,k,p,q=1}^m{\bra{j}U_x\ket{k}}^*\bra{j}U_y\ket{k}\bra{p}U_x\ket{q}{\bra{p}U_y\ket{q}}^* \\
&=& \frac{1}{m^2}\sum_{x,y=1}^n\sum_{j,k,p,q=1}^m\exp\left[\frac{2\pi i \big(p-j+m(q-k)\big)(x-y)}{n}\right] \\
&=& \frac{n^2}{m^2}\sum_{j,k,p,q=1}^m\delta_{jp}\delta_{qk}\;=\;n^2\;
\end{eqnarray}
provided $n\geq m^2$.

To achieve the above worst-case error probability [Eq. (\ref{Qwcep})], Alice and Bob first convert 
the maximally entangled state $\ket{\psi_+^{(n!)}}$ into a uniformly distributed shared random variable 
$\xi\in\{1,..,n!\}$ through local measurements in the computational basis. They now use $\xi$ to jointly choose $\pi_\xi$, one of
$n!$ different random permutations of Sapna's message labels. The second maximally entangled state,
$\ket{\psi_+^{(m)}}$, is used in manner similar to Theorem \ref{theorem5}. Alice and Bob perform the local
operation $U_{\pi_\xi(x)}^*\otimes U_{\pi_\xi(y)}$ to $\ket{\psi_+^{(m)}}$, where $U_x$ is now defined as in
Eq. (\ref{UTOF}), and send the result to Roger. Roger performs the projective measurement
$\big\{P_1=\ket{\psi_+^{(m)}}\bra{\psi_+^{(m)}}, P_0=1-P_1\big\}$, revealing result 1 with probability
\begin{equation}
\frac{1}{n^2-n}\sum_{x\neq y}\left|\bra{\psi_+^{(m)}}U_{\pi_\xi(x)}^*\otimes U_{\pi_\xi(y)}\ket{\psi_+^{(m)}}\right|^2
\;=\; \frac{1}{n^2-n}\left(\frac{1}{m^2}\sum_{x,y}\left|\tr\left[U_x^{\dag}U_y\right]\right|^2-n\right)\;=\;\frac{n^2/m^2-n}{n^2-n}
\end{equation}
when $x\neq y$, and result 1 with probability 
\begin{equation}
\frac{1}{n}\sum_{x}\left|\bra{\psi_+^{(m)}}U_{\pi_\xi(x)}^*\otimes U_{\pi_\xi(x)}\ket{\psi_+^{(m)}}\right|^2
\;=\; \frac{1}{n}\left(\frac{1}{m^2}\sum_{x}\left|\tr\left[U_x^{\dag}U_x\right]\right|^2\right)\;=\;1
\end{equation}
when $x=y$. Thus, the protocol has one-sided error and a worst-case error
probability given by Eq. (\ref{Qwcep}).
\end{proof}

\section{Conclusion}

To summarize, we have derived the minimum achievable worst-case error probability for classical fingerprinting protocols
with one-sided error and an arbitrary amount of shared randomness. This is our main result and the content of Theorem \ref{theorem4}.
Furthermore, we have presented entanglement-assisted quantum fingerprinting protocols (Theorems \ref{theorem5} and \ref{theorem6}) with error
rates surpassing the best classical protocols. We hope that our work provides some important new results applicable to current
experimental investigations of quantum fingerprinting protocols \cite{Horn04,Du04}.

Our analysis is by no means complete. Future research directions might include: deriving the minimum achievable worst-case
error probability for entanglement-assisted quantum fingerprinting protocols, investigating the required amount of
shared randomness/entanglement necessary to execute fingerprinting protocols, or deriving error bounds for fingerprinting
protocols with two-sided error.

The absolute limits of successful fingerprinting protocols provide quantitative measures for the compressibility of 
information stored in message strings. Our analysis may be appended to the growing list which reveal a fundamentally 
greater capacity to compress data stored as quantum information.

\begin{acknowledgments}
This work has been supported by CIAR, CSE, iCORE and MITACS. JW acknowledges
support from AIF and PIMS.
\end{acknowledgments}

\end{document}